\documentclass[12pt]{article}

\usepackage{graphicx}
\usepackage{xcolor}
\usepackage{physics}
\usepackage{bm}
\usepackage{amsfonts}
\usepackage{scicite}
\usepackage{times}
\newenvironment{sciabstract}{%
\begin{quote} \bf}
{\end{quote}}

\title{Non-Hermitian delocalization\\ in a 2D photonic quasicrystal}

\author{Zhaoyang Zhang$^{1\ast}$, Shun Liang$^1$, Ismaël Septembre$^2$, Jiawei Yu$^1$,\\ Yongping Huang$^1$, Maochang Liu$^3$, Yanpeng Zhang$^1$,\\  Min Xiao$^{4,5}$, Guillaume Malpuech$^{2\ast}$, Dmitry Solnyshkov$^{2,6\ast}$\\
\\
\normalsize{$^{1}$Key Laboratory for Physical Electronics and Devices of the Ministry of Education}\\
\normalsize{\& Shaanxi Key Lab of Information Photonic Technique,}\\ 
\normalsize{School of Electronic Science and Engineering,}\\
\normalsize{Faculty of Electronics and Information, Xi'an Jiaotong University, Xi'an 710049, China}\\
\normalsize{$^2$Institut Pascal, PHOTON-N2, Universit\'e Clermont Auvergne, CNRS, Clermont INP,}\\
\normalsize{F-63000 Clermont-Ferrand, France}\\
\normalsize{$^3$International Research Center for Renewable Energy}\\
\normalsize{\& State Key Laboratory of Multiphase Flow in Power Engineering,}\\
\normalsize{Xi'an Jiaotong University, Xi'an 710049, China}\\
\normalsize{$^4$Department of Physics, University of Arkansas, Fayetteville, Arkansas, 72701, USA}\\
\normalsize{$^5$National Laboratory of Solid State Microstructures and School of Physics,}\\
\normalsize{Nanjing University, Nanjing 210093, China}\\
\normalsize{$^6$Institut Universitaire de France (IUF), 75231 Paris, France}\\
\\
\normalsize{$^\ast$To whom correspondence should be addressed;}\\
\normalsize{E-mails:  zhyzhang@xjtu.edu.cn, guillaume.malpuech@uca.fr, dmitry.solnyshkov@uca.fr.}
}
\date{}
\begin{document}
\baselineskip24pt


\maketitle


\begin{sciabstract}
Quasicrystals show long-range order, but lack translational symmetry. So far, theoretical and experimental studies suggest that both Hermitian and non-Hermitian quasicrystals show localized eigenstates. This localization is due to the fractal structure of the spectrum in the Hermitian case and to the transition to diffusive bands via exceptional points in the non-Hermitian case. Here, we present an experimental study of a dodecagonal (12-fold) photonic quasicrystal based on electromagnetically-induced transparency in a Rb vapor cell. The transition to a quasicrystal is obtained by superposing two honeycomb lattices at 30$^\circ$ with a continuous tuning of their amplitudes. Non-Hermiticity is controlled independently. We study the spatial expansion of a probe wavepacket. In the Hermitian case, the wavepacket expansion is suppressed when the amplitude of the second lattice is increased (quasicrystal localization). We find a new regime, where increasing the non-Hermitian potential in the quasicrystal enhances spatial expansion, with the $C_{12}$ symmetry becoming visible in the wavepacket structure. This real-space expansion is due to a k-space localization on specific quasicrystal modes. Our results show that the non-Hermitian quasicrystal behavior is richer than previously thought. The localization properties of the quasicrystals can be used for beam tailoring in photonics, but are also important in other fields.
\end{sciabstract}

Teaser: A combination of two localizing mechanisms leads to delocalization in non-Hermitian photonic quasicrystals.

\section*{Introduction}
Quasicrystals are characterized by long-range order without translational symmetry\cite{Levine1984}. In mathematics, they correspond to infinite non-periodic tilings. They can possess rotational symmetries incompatible with the translational one, such as the famous pentagonal symmetry of the Penrose tiling\cite{Shechtman1984}. Another interesting and important case is the dodecagonal symmetry\cite{Ishimasa1985,yang1987description,Gahler1988}, which can be obtained from a superposition of two honeycomb lattices\cite{niizeki1987two,Zhang2001,ahn2018dirac,crosse2021trigonal} rotated by 30$^\circ$. This configuration is particularly timely, because of the extreme popularity of moiré honeycomb lattices, such as magic angle twisted bilayer graphene\cite{andrei2020graphene}, obtained for angles of rotation smaller than $30^\circ$. Moiré lattices and quasicrystals share many common properties, such as the presence of flat bands\cite{Tarnopolsky2019,Moon2019,Huang2019} in their spectrum. Dodecagonal quasicrystals are studied in many fields: chemistry\cite{Hayashida2007,zhang2012dodecagonal,gillard2016dodecagonal,jayaraman2021dodecagonal}, material science\cite{yang1987description,fischer2011colloidal,xiao2012dodecagonal}, electronics\cite{ahn2018dirac}, topological physics\cite{Kraus2013,Tran2015,Hua2020}, and photonics\cite{chan1998photonic,kaliteevski2000two,zoorob2000complete,Zhang2001,Feng2005,man2005experimental,gauthier2005photonic,nozaki2006lasing,Ren2018,Xi2019}.

For 1D quasicrystals or quasiperiodic lattices, many important analytical results were obtained using the Aubry-André model\cite{aubry1980analyticity}: instead of considering a structure without translational symmetry in the positions of individual sites, one considers a periodic lattice with an incommensurate on-site potential of a variable strength\cite{Lahini2009,Ganeshan2013,dominguez2019aubry,Longhi2019}. It is now theoretically established and experimentally demonstrated that the dispersion of such a 1D quasicrystal contains an infinite number of gaps which obey the gap labeling theorem\cite{simon1982almost,bellissard1992gap,gambaudo2014brillouin,Tanese2014fractal}. Each single band is infinitely narrow (flat), and the mobility of the particles filling the bands is strongly suppressed\cite{roati2008anderson,Lahini2009}. This model allows studying the transition towards the fractal energy spectrum and the associated localization \cite{goblot2020emergence}, driven by the variable strength of the on-site potential.

2D quasicrystals have also been studied theoretically using the Aubry-André approach \cite{Huang2019}, namely considering a superposition of two lattices in 2D: one lattice is fixed, while the strength of the second is varied, allowing to observe the modification of the transport. Another theoretical approach was to start directly with a quasicrystal potential and vary its strength relative to the recoil energy\cite{ueda1987energy,Szabo2020,Gautier2021,zhu2023localization}, allowing to see the localization of some of the eigenstates described by their inverse participation ratio. The bands were shown to tend to a Cantor set analogue \cite{damanik2011spectral}, as in 1D \cite{sutHo1989singular}. In experiments with Hermitian 2D quasicrystals, phononic\cite{he1989eigenvalue} and photonic \cite{vardeny2013optics} bandgaps were explicitly observed, in particular in dodecagonal structures\cite{zoorob2000complete}. We also note a recent demonstration of enhancement of the transport by disorder\cite{levi2011disorder}.

The potential can also be imaginary, making possible non-Hermitian phenomena analogous to the PT-symmetry-breaking transition, well-known in modern photonics \cite{Ozdemir2019}. Such transition has recently been predicted\cite{Longhi2019} and observed experimentally\cite{Lin2022} in a 1D quasicrystal: increasing the non-Hermiticity induces a phase transition, which ultimately suppresses the mobility edge. All states become localized, and the mechanism is not due anymore to the quasi-crystal flat bands, but to the emergence of diffusive non-Hermitian bands (Fermi arcs limited by exceptional points). The Aubry-André approach has often been used for non-Hermitian systems \cite{Zeng2020,Li2022}. Theoretical analyses of 2D systems have also been performed, based on a specific complex potential case\cite{xu2022exact}, similar to the one considered in the 80s\cite{sarnak1982spectral} and showing qualitatively similar results with respect to the 1D case.

In this work, we take advantage of a reconfigurable photonic platform, atomic vapors under electromagnetically-induced transparency (EIT)\cite{PhysRevA.51.576} in a three-level atomic configuration\cite{Zhang2019,zhang2020spin,zhang2022imaging}, to perform an experimental study of a 2D Hermitian and non-Hermitian dodecagonal quasicrystals with a tunable ratio of intensities between the two honeycomb lattices forming the quasicrystal and a separately tunable non-Hermiticity. We demonstrate the localization transition with the increase of the lattice ratio in the Hermitian case. On the contrary, in the non-Hermitian case the initial localization is followed by a delocalization. The latter is caused by the wavepacket redistribution due to the lifetime difference, occurring without crossing exceptional points.

\section*{Results}

\begin{figure}
\centering
\includegraphics[width=0.8\linewidth]{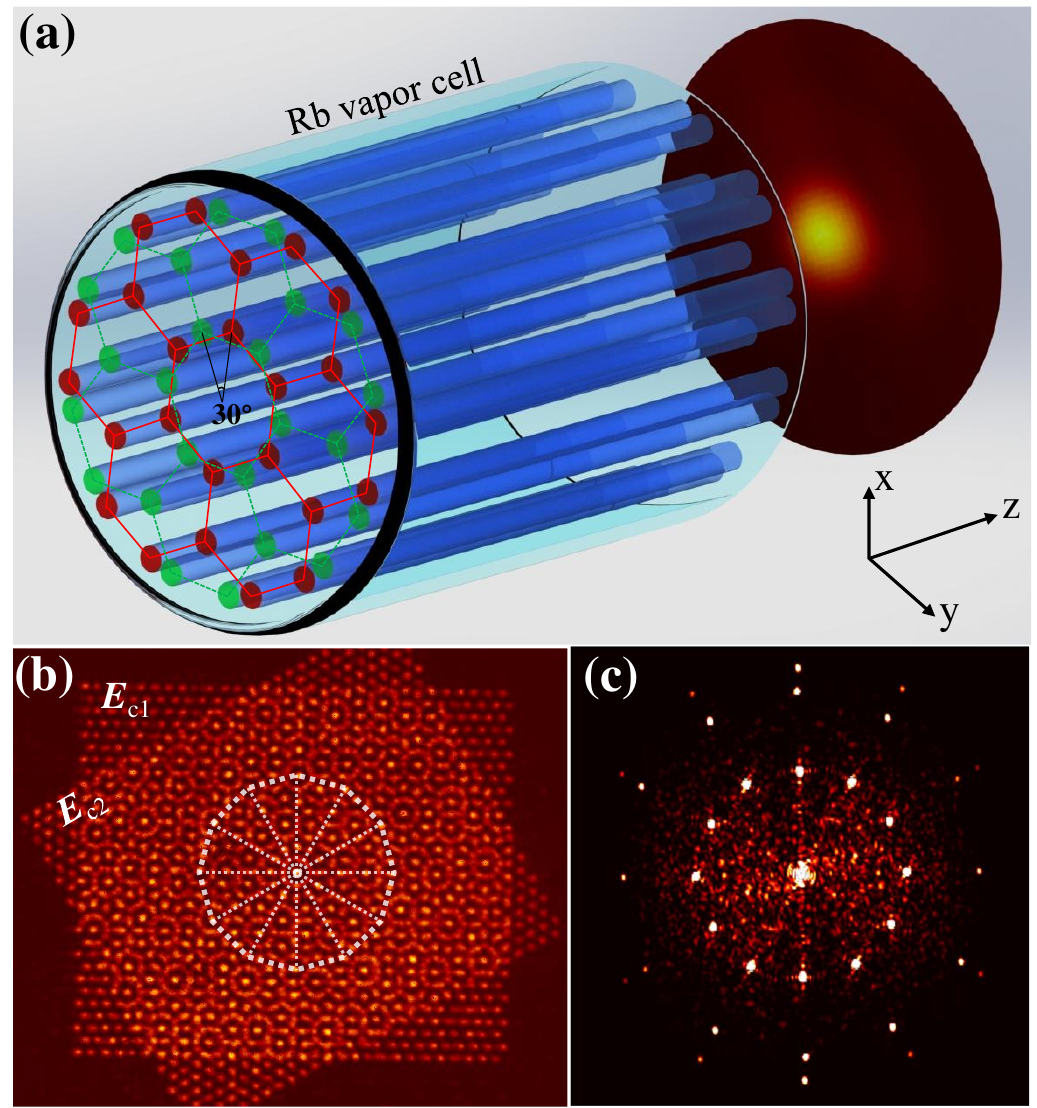}
\caption{\textbf{Experimental configuration and the real and reciprocal space images of the quasicrystal.} 
\textbf{a} Experimental scheme. \textbf{b} The experimentally generated dodecagonal quasicrystal lattice formed by two hexagonal patterns rotated by $30^\circ$. \textbf{c} Reciprocal-space image of the experimental quasicrystal lattice exhibiting a 12-fold symmetry in 3 orders of diffraction. \label{fig1}}
\end{figure}

The experimental scheme is shown in Fig.~\ref{fig1}a.  Two honeycomb photonic lattices are optically induced inside a Rb vapor cell by two hexagonal coupling beams \textbf{\textit E}$_{C1}$
(frequency $\omega_{c1}$) and \textbf{\textit E}$_{C2}$ ($\omega_{c2}$) with the same period of 200 $\rm{\mu m}$. Both coupling beams are injected into the vapor cell along the $z$ direction. There exists a rotation angle (in the $x-y$ plane) of $30^{\circ}$ between the two hexagonal patterns generated by a phase-controlled spatial light modulator (SLM). A weak Gaussian probe beam \textbf{\textit E}$_p$ ($\omega_{p}$) from a continuous-wave tunable external cavity diode laser (ECDL) co-propagates with coupling beams to excite a three-level atomic configuration (see Methods for a scheme), where the well-known EIT effect can occur at appropriate detunings satisfying the two-photon resonance\cite{PhysRevA.51.576} $\delta_{p}-\delta_{c1}(\delta_{c2})=0$. Here, the frequency detunings $\delta_i$ ($i=p$, $c1$ and $c2$) are defined as the difference between the energy gap between the levels driven by laser field \textbf{\textit E}$_i$ and its frequency (see Methods for more details on the experimental setup). Under the EIT condition, the susceptibility $\chi$ experienced by \textbf{\textit E}$_p$ is inversely related to the intensity of the coupling beams\cite{Zhang2020,PhysRevLett.131.013802}.
This intensity is shown in Fig.~\ref{fig1}b. Each coupling beam forms a single honeycomb photonic lattice, corresponding to the dark sites of a hexagonal pattern visible in the figure. The propagation of a probe beam through the vapor cell with an EIT-induced susceptibility distribution is described by the paraxial equation:
\begin{equation}
    i\frac{\partial E}{\partial z}=-\frac{1}{2k_0}\Delta E-\frac{k_0\chi}{2}E\text{,}
\end{equation}
where $k_0$ is the probe wave vector. This is equivalent to a 2D time-dependent Schr\"odinger equation with $z\sim t$ (time), $k_0\sim m$ (particle mass), and $\chi\sim -U$ (external potential). Susceptibility maxima (dark sites in Fig.~\ref{fig1}b) thus correspond to potential minima.

The transmitted probe beam is received by a charge-coupled device (CCD) camera (placed behind the output plane of the cell) through an imaging lens. During the experiment, the detuning of the probe beam is set as $\delta_p=-260$ MHz, while $\delta_{c1}$ and $\delta_{c2}$ are manipulated [around positive two-photon detuning $\delta_{p}-\delta_{c1}(\delta_{c2})$] as required to control the degree of non-Hermiticity of the induced photonic lattice (see  Methods for values of the detunings used for each figure). 
The 12-fold symmetry of the resulting lattice is underlined in Fig.~\ref{fig1}b by the white dodecagon. Figure~\ref{fig1}c shows the reciprocal-space image also exhibiting a 12-fold pattern with the first three orders of diffraction clearly visible, which confirms the formation of a quasicrystal \cite{ahn2018dirac}.

We now study the evolution of the probe beam in the quasicrystal potential created by the coupling beams. 
The probe beam represents a narrow wavepacket, equivalent to the excitation of the vicinity of a single lattice site. The duration of the time evolution in the 2D Schr\"odinger equation is fixed by the length of the vapor cell in the $z$ direction. It is sufficient for the wavepacket to expand over several unit cells in a honeycomb lattice, whereas in the quasicrystal configuration the expansion is expected to be suppressed.

\begin{figure}
\centering
\includegraphics[width=0.8\linewidth]{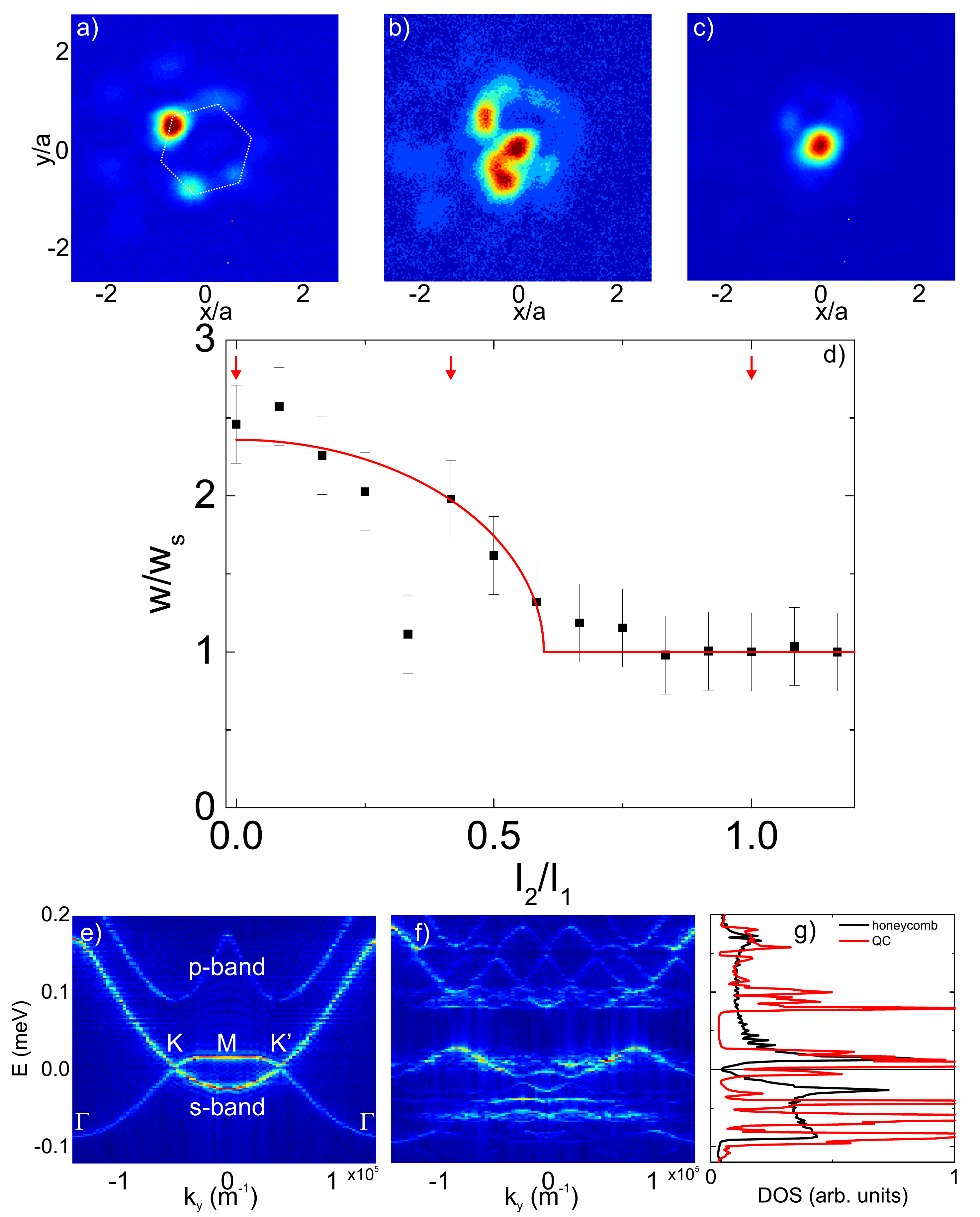}
\caption{\textbf{Wavepacket expansion and the localization transition with the increase of the second lattice strength.} \textbf{a}-\textbf{c} Spatial images of the wavepacket after its evolution in the Hermitian lattice (lattice intensity ratio $I_2/I_1=0$, $0.4$, $1$, respectively). \textbf{d} Wavepacket width $w$ normalized by the 1-site width $w_s$. Red arrows mark the correspondence with panels a-c. 
\textbf{e} The dispersion of a single honeycomb lattice through $\Gamma K M K' \Gamma'$ points. \textbf{f} The dispersion of a quasicrystal showing multiple gaps. \textbf{g} The comparison of the DOS for a periodic honeycomb lattice and a quasicrystal. The gaps appear as zeroes of the DOS. \label{fig2}}
\end{figure}

Figure~\ref{fig2} presents the results obtained in the fully Hermitian case. We keep one honeycomb lattice turned on with a constant power $I_1$, while varying the intensity of the second lattice $I_2$. The top panels Fig.~\ref{fig2}a-c show the spatial distribution of the output probe patterns for three ratios of $I_2/I_1$ ($0$, $0.4$, $1$, respectively). A clear narrowing of the output wavepacket can be observed. We have systematically studied the width of the output wavepacket as a function of the ratio $I_2/I_1$. The results are shown in Fig.~\ref{fig2}d with black dots with error bars corresponding to the uncertainty of the extraction.

The output width of the wavepacket exhibits a continuous decrease until it drops to its minimal size, approximately corresponding to the size of a single lattice site $w_s$ that we take as a reference for this plot. To explain this behavior and to determine the transition point, we have performed numerical simulations based on the paraxial approximation (see Methods for details). An example of the dispersion of a single honeycomb lattice that we use as a starting point of our analysis is shown in Fig.~\ref{fig2}e. It is plotted along the $\Gamma K M K'\Gamma$ high-symmetry points. 
As expected from the theory of incommensurate potentials and quasicrystals, the increase of $I_2/I_1$ up to $1$ opens a set of gaps in the dispersion, making the band similar to a Cantor set. A second example of the dispersion for $I_2/I_1=1$ is shown in Fig.~\ref{fig2}f. It indeed exhibits a lot of gaps separating bands which become very narrow. The gaps and the bands in 2D can be efficiently analyzed via the density of states (DoS), allowing us to observe full gaps and to determine their size. Figure~\ref{fig2}g shows the DoS for the two cases shown in panels e and f: honeycomb lattice and dodecagonal quasicrystal. The Dirac point is visible for the honeycomb lattice (black) as a zero-DOS point with linear behavior in its vicinity. In the quasicrystal case (red), multiple large gaps accompanied by narrower secondary gaps are visible. The edges of each gap demonstrate van Hove singularities (DoS peaks corresponding to band edges).

In general, the wavepacket expansion is determined by the group velocity of its components. If the wavepacket is narrow in real space, it covers the whole Brillouin zone and thus allows to probe the \emph{maximal} group velocity available. Our simulations show that the first (and largest) gap is opened precisely at the point of highest group velocity, because here the wavefunction is the most sensitive to the perturbing potential. It corresponds to the $\Gamma K$ direction, where the dispersion of a single honeycomb lattice is given by $E(k)=\pm J(1+2\cos ka/2)$ in the tight-binding limit, and the group velocity is $v_g(k)=\pm\hbar^{-1}a \sin ka/2$, with the maximal $v_g$ point $k_{max}=\pi/a$. The gap size $\Delta$ is linearly proportional to the strength of the incommensurate potential $\lambda= I_2/I_1$ for small perturbations: $\Delta\sim\lambda$. This allows to find the behavior of the wavepacket expansion via the group velocity as a function of the perturbation strength $\lambda$:
\begin{equation}
    \frac{w(I_2/I_1)}{w_s}=1+A \sqrt{1-B\left(\frac{I_2}{I_1}\right)^2}\text{,}
    \label{widthfuction}
\end{equation}
where $A$ is the proportionality coefficient between the group velocity and the wavepacket width (including the effective propagation time), while $B$ is the proportionality coefficient between the gap size $\Delta$ and the perturbation $\lambda$. The red curve in Fig.~\ref{fig2}d fits the experimental data with Eq.~\eqref{widthfuction}, giving the fitting parameters $A\approx 1.35\pm 0.09$ and $B\approx 2.81\pm 0.12$. This allows to obtain the localization transition point $(I_2/I_1)_{loc}=\sqrt{1/B}\approx 0.597\pm 0.013$, of the same order of magnitude as in other quasicrystals. We therefore conclude that we have observed a localization transition for a Hermitian 2D dodecagonal quasicrystal and found its approximate position. Of course, the transition point depends on the properties of the periodic potential, which in our case is relatively weak (as can be seen from the shape of the $p$-band visible in Fig.~\ref{fig2}e).

The special localization point visible in Fig.~\ref{fig2}d for $I_2/I_0\approx 0.3$ occurs because some localized states appear in a perturbed periodic potential even before the opening of multiple gaps due to the fractalization of the energy spectrum. If the probe has a strong overlap with such state, the wavepacket will not expand, even though there are still propagative states available in some bands: they are simply not excited efficiently. The recovery of the expansion with the increase of the second lattice potential $I_2$ confirms the "occasional" nature of this localization. On the contrary, the expansion is definitively suppressed once the true localization threshold is passed.

We now turn to the non-Hermitian case. Indeed, the EIT configuration allows varying not only the real part of the effective potential controlled by the susceptibility, but also the imaginary part of the susceptibility, potentially providing an important non-Hermiticity to the potential. It ultimately allows to observe a transition similar to the PT-symmetry-breaking one\cite{zhang2022imaging}, but in the present work we remain below this transition, defined by a critical value of $(\chi''/\chi')_{crit}\approx 0.4$ (here we use $(\chi''/\chi')\approx 0.2$). As in the Hermitian case, we fix the intensity of the first honeycomb lattice $I_1$ and vary the intensity of the other $I_2$, with both lattices being non-Hermitian. We note that the real part of the potential is different from that of Fig.~\ref{fig2}a-d.

\begin{figure}
\centering
\includegraphics[width=0.8\linewidth]{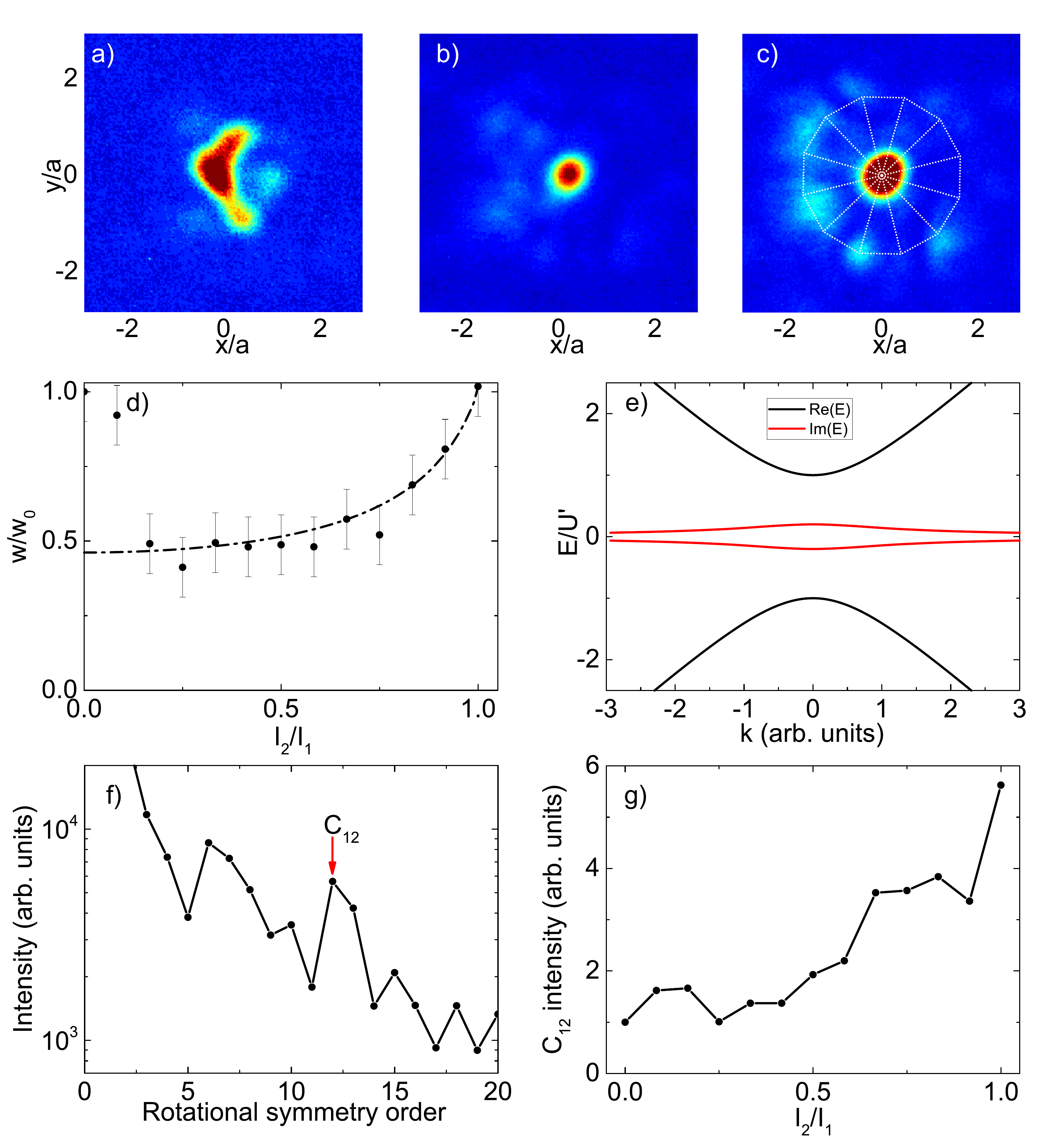}
\caption{\textbf{localization-delocalization transition in a 2D non-Hermitian quasicrystal.} 
\textbf{a}-\textbf{c} Spatial images of the wavepacket after its evolution in the non-Hermitian lattice (lattice intensity ratio $I_2/I_1=0.1$, $0.4$, $1$, respectively). \textbf{d} Wavepacket width $w$ normalized by the reference width $w_0$ (corresponding to $I_2/I_1=0$). Points with error bars (instrumental uncertainty) -- experiment, dash-dotted line -- theory.
\textbf{e} Real (black) and imaginary (red) parts of the eigenenergies of the weak complex potential model.
\textbf{f} Fourier-transform of the angular pattern of the panel c ($I_2/I_1=1$) exhibiting a maximum corresponding to dodecagonal symmetry $C_{12}$. \textbf{g} Intensity of the $C_{12}$ maximum of the Fourier transform as a function of $I_2/I_1$: the symmetry of the wavepacket inherits that of the lattice.
\label{fig3}}
\end{figure}

Figure~\ref{fig3}a-c shows the spatial images of the output beam for three values of $I_2/I_1$ ($0.1$, $0.4$, and $1$, respectively). Interestingly, after the onset of localization, the wavepacket expansion is recovered almost completely, and the symmetry of the final wavepacket state changes. Figure~\ref{fig3}d shows the wavepacket size $w$ (black dots) normalized by the size $w_0$ observed for a single honeycomb lattice $I_2/I_1=0$. The measurements demonstrate a minimum around $I_2/I_1\approx 0.4$.

To understand this behavior, we use the weak potential approximation and work with an effective Hamiltonian (see Methods for details). This allows us to obtain the asymptote shown in Fig.~\ref{fig3}d with a black dash-dotted line. It describes the wavepacket broadening due to the non-Hermitian mechanism described by the following Hamiltonian:
\begin{equation}
\label{HamNH}
    H=\alpha (k-k_0) \sigma_z+U'\sigma_x+iU''\sigma_x\text{.}
\end{equation}
This Hamiltonian exhibits exceptional points if $U'=0$, that is, if the potential is purely imaginary. The position of exceptional points is determined by $(k^*-k_0)=\pm U''/\alpha$. In our case, these points are not accessible, since $U'\neq 0$.  We actually have $U''=\eta U'$ with $\eta=0.2$. Nevertheless, the non-Hermitian nature of the Hamiltonian leads to important consequences: the decay rate of the states (the imaginary part of the energy) starts to depend on their wave vector.
The eigenvalues are given by $E(k-k_0)=\pm\sqrt{U'^2(1-i\lambda)^2+\alpha^2 (k-k_0)^2}$. The maximum of the imaginary part is $\gamma=\pm\lambda U'$, while the width of the decay rate distribution is $\sigma_\gamma=U'\alpha^{-1}$ (see Fig.~\ref{fig3}e showing the corresponding correction to the overall decay rate). The resulting decay rate profile leads to the concentration of the wavepacket at longest-living states in the reciprocal space at the edge of the largest gap, corresponding to real-space expansion. The wavepacket width grows as a function of the ratio of the two lattices for fixed evolution time $t$, according to the following law (see Methods): 
\begin{equation}
    \Delta r=\frac{A}{\sqrt{1-B\left(\frac{I_2}{I_1}\right)^2}}\text{.}
\end{equation}
Fig.~\ref{fig3}d shows a fitting with $A\approx 0.46\pm 0.02$ (consistent with Fig.~\ref{fig2}: the wavepacket expansion gives a factor $A^{-1}\approx 2$ with respect to a single site) and $B\approx 0.79 \pm 0.02$ (meaning that absorption length due to the non-Hermiticity is shorter than the vapor cell length). The theoretical curve presents a good agreement with the experimental data. We therefore conclude that while in periodic systems the non-Hermiticity can lead to localization via the PT-symmetry-breaking transition, in our quasicrystal we observe that the non-Hermiticity leads to delocalization in wavepacket expansion. We note that delocalization has been observed in pentagonal quasicrystals\cite{levi2011disorder}, but there it was induced by disorder and not by non-Hermiticity.

Contrary to the Hermitian case, where the wavepacket localization width is comparable to the size of a single lattice site $w_s$, the non-Hermitian case, thanks to the suppression of the localization, allows to observe the wavepacket distribution over several neighboring sites for $I_2/I_1=1$ (exact quasicrystal limit). We analyze the angular distribution of this wavepacket (the probability density averaged over the radial coordinate $r$) by performing its Fourier transform shown in Fig.~\ref{fig3}f. A clear maximum corresponding to the dodecagonal ($C_{12}$) symmetry is observed. The corresponding dodecagon is marked in Fig.~\ref{fig3}c with white dashed lines. This confirms that the wavepacket inherits the symmetry of the quasicrystal lattice. We also study the behavior of the $C_{12}$ maximum of the angular Fourier transform with the intensity of the second lattice $I_2/I_1$ in Fig.~\ref{fig3}g (normalized to its "background" value at $I_2/I_1=0$) and observe a strong growth of this component above $I_2/I_1\approx 0.6$, when the wavepacket delocalization also takes place. This confirms that for small intensity of the second lattice its effect can be seen as an incommensurate (effectively random) on-site potential for the initial (honeycomb) lattice, whereas for large intensities the superposition of two lattices must be indeed considered as a dodecagonal quasicrystal with associated properties.

\section*{Discussion}

Quasicrystals now fascinate scientists not only by the mere fact of their existence, but also by their properties. In photonics, quasicrystals are easy to implement and provide a wide range of applications for beam shaping. We have studied the beam evolution in a reconfigurable photonic platform, allowing us to continuously analyze the transition between a crystal and a quasicrystal both in Hermitian and non-Hermitial cases. We have observed an efficient localization of the beam in Hermitian quasicrystals. We have also shown that the combination of two localizing contributions (incommensurate potential and non-Hermiticity) can actually lead to delocalization, allowing us to recover almost the same transport properties as in the Hermitian case, but with the wavepacket symmetry becoming dodecagonal.

The main limitation of all experimental methods of the studies of quasicrystals, especially in analogue systems, is the finite size of the lattice. It prevents the observation of high-order diffraction peaks and gaps, as well as the observation of strictly flat bands. The localization transition is necessarily broadened. Other finite-size constraints include the length of the vapor cell, which does not allow the observation of the wavepacket expansion over very large distances. In the non-Hermitian case, the losses weaken the signal and decrease the precision of measurements. Because of this, the recovery of the wavepacket expansion is combined with the attenuation of the probe beam intensity.

Our work can find direct applications for on-demand beam tailoring\cite{miyai2006lasers,perez2017demand,fu2020optical}. Generally speaking, the applications of quasicrystals in photonics go beyond the localization \cite{vardeny2013optics}, waveguiding\cite{jin1999band} and beam focusing \cite{di2008parametric}: in particular, they were also shown to exhibit negative refraction\cite{Feng2005}.

\section*{Materials and Methods}
\textbf{Experimental setup.} The two hexagonal coupling beams \textbf{\textit E}$_{C1}$ and \textbf{\textit E}$_{C2}$ emit from the left and right half of the screen of the liquid crystal SLM (the resolution is 1920 × 1152; loaded with a 256-bit phase hologram), while its two incident Gaussian beams are from two semiconductor tapered amplifiers (TAs), respectively. The input beams of TAs are respectively derived from two ECDLs different from that of \textbf{\textit E}$_p$. The wavelength of the three beams is around 795.0 nm. The power of the probe beam is 270 $\mu$W. The available maximum power of a single hexagonal coupling beam is 45 mW. The left and right half of the SLM’s screen are set as two independent regions and loaded with different phase diagrams (according to the weighted Gerchberg-Saxton algorithm) to create the required coupling fields. The probe beam is horizontally polarized while the coupling beams are vertically polarized. This polarization arrangement makes only the transmitted probe beam enter the CCD camera while the coupling beams can be easily filtered out by a polarization beam splitter. The 2.5~cm long Rb atomic vapor cell is wrapped with $\mu$-metal sheets to shield the environmental magnetic field and heated to 100$^\circ$C by a home-made temperature controller. \\
Experimental images obtained from the CCD camera are then analyzed to extract the wavepacket size shown in Figs.~\ref{fig2} and \ref{fig3}. The noise contribution is removed before the calculation of the root mean square width of the wave packet $w=\sqrt{\langle(\bm{r}-\langle\bm{r}\rangle)^2\rangle}$.\\
\textbf{Susceptibility.} Under the EIT condition, the susceptibility experienced by \textbf{\textit E}$_p$ is described as:
\begin{equation}
   \chi  = \frac{{iN{{\left| {{\mu _{31}}} \right|}^2}}}{{\hbar {\varepsilon _0}}} \times {\left( {({\Gamma _{31}} - i{\delta _p}) + \frac{{|{\Omega _{c1}}{|^2}}}{{{\Gamma _{32}} - i({\delta _p} - {\delta _{c1}})}} + \frac{{|{\Omega _{c2}}{|^2}}}{{{\Gamma _{32}} - i({\delta _p} - {\delta _{c2}})}}} \right)^{ - 1}}
    \label{susceptibility}
\end{equation}
where $N$ is the atomic density, $\mu_{mn} (m, n$=1,2 and 3) and $\Gamma_{mn}$ are the respective dipole moment and decay rate between levels $|m\rangle$ and $|n\rangle$ connected by corresponding beams, $\Omega_i$ is the Rabi frequency of the laser field \textbf{\textit E}$_i$, and is directly proportional to its electric field intensity. The corresponding energy-level diagram of generating EIT is given in the Supplementary Materials (Figure~S1). By properly setting the detuning of either coupling beam, it can establish a honeycomb photonic lattice with different degrees of non-Hermiticity.\\
\textbf{Detunings.} We have used the following values of the detunings for the figures. Figure~\ref{fig2} (Hermitian case): $\Delta_p=-260$~MHz, $\Delta_c=-290$~MHz, the real part modulation is estimated as $\chi'\approx 8.8\times 10^{-4}$, and the imaginary part is $\chi''\approx 2.07\times 10^{-5}$. The ratio of imaginary and real susceptibilities is  $\chi''/\chi'\approx 0.02$. Figure~\ref{fig3} (non-Hermitian case):  $\Delta_p=-30$~MHz, $\Delta_c=-40$~MHz, and the real part is estimated as $\chi'\approx 1.4\times 10^{-3}$, while the imaginary part is $\chi''\approx 2.9\times 10^{-4}$. The ratio of imaginary and real susceptibilities is $\chi''/\chi'\approx 0.207$. Supplementary Materials present additional results obtained at different detunings indicated in the Supplementary Text.

\noindent{\textbf{Theory.}} We have performed numerical simulations in the paraxial approximation. To solve the equation~(1) written in the main text, we used the combination of the 3rd order Adams-Bashforth method with Fourier-transform calculation of the Laplacian operator accelerated by the Graphics Processor Unit. The real and imaginary parts of the susceptibility were calculated using Eq.~\eqref{susceptibility}. The dispersions shown in Fig.~\ref{fig2} were calculated using a narrow probe excitation (smaller than a single lattice site). The resulting solution $E(x,y,z)$ was Fourier-transformed over all coordinates to obtain the dispersion $|E(k_x,k_y,\hbar\omega)|^2$ (with $\hbar\omega=\hbar c k_z$), whose cuts are shown in Fig.~\ref{fig2}e,f. 
The probe was introduced as an initial condition ($z=0$) for the complex electric field $E$. The density of states was obtained by integrating the resulting probability density $|E(\omega,k_x,k_y)|^2$ over all in-plane directions. 

To describe the effect of the imaginary part of the potential, we use the weak potential approximation, whose validity is justified by the examples of numerically calculated dispersion shown in Fig.~\ref{fig2}e, exhibiting a strong mixing between the $s$ and $p$ bands. In this approximation, the coupling of the states $\mathbf{k}$ and $\mathbf{k}'$ due to a  potential $U(\mathbf{r})$ is described by the following Hamiltonian:
\begin{equation}
H = \left( {\begin{array}{*{20}{c}}
{E\left( {\bf{k}} \right)}&{{U_{{\bf{kk}}'}}}\\
{{U_{{\bf{k}}'{\bf{k}}}}}&{E\left( {{\bf{k}}'} \right)}
\end{array}} \right)
\label{nearlyfree}
\end{equation}
where
\begin{equation}
    U_{\mathbf{kk}'}=\int \mathbf{dr}\, U(\mathbf{r})e^{i(\mathbf{k}-\mathbf{k'})\mathbf{r}}
\end{equation}
This function depends only on the difference $\mathbf{k}-\mathbf{k}'$. Therefore, $U(\mathbf{k}-\mathbf{k}')$ represents the Fourier transform of the potential, whose experimental image for the case $I_2/I_1=1$ is shown in Fig.~\ref{fig1}c. 
As the intensity of the second lattice is increased from zero, the maxima corresponding to the $C_{12}$ symmetry also increase linearly. The only non-zero elements of $U(\mathbf{k}-\mathbf{k}')$ correspond to these peaks. This fixes the difference between $\mathbf{k}$ and $\mathbf{k}'$, and we can therefore keep only one of these two wave vectors as a parameter of the Hamiltonian. 

We focus on the region of the maximal group velocity of the first lattice, where the largest gap is opened. It corresponds to the first-order diffraction maximum of the quasicrystal lattice. A similar analysis can be applied to all other orders. This region corresponds to the inflection point of the dispersion, where the energy depends linearly on the wave vector: $E(k)=\alpha (k-k_0)$, and $E(k')=E(k-g)=-\alpha (k-k_0)$. The direction of $k$ is aligned here with the direction from $k=0$ to any of the 1st-order diffraction maxima of the quasicrystal lattice.

We choose the origin of the coordinates in order for the potential to be an even function of coordinates (both its real and imaginary parts $U'$ and $U''$). Due to the properties of the Fourier transform, $U(\mathbf{k}-\mathbf{k}')=U'(\mathbf{k}-\mathbf{k}')+iU''(\mathbf{k}-\mathbf{k}')$ is also an even function. The Hamiltonian therefore becomes non-Hermitian. Its expression is given in the main text (Eq.~\eqref{HamNH}). We now describe the evolution of the wavepacket distribution in the reciprocal space. Initially, the wavepacket is a Gaussian centered at $k=0$: $|\psi(k)|^2=\exp(-k^2/(2\sigma_0^2))/\sqrt{2\pi\sigma_0^2}$. The amplitudes of its components evolve exponentially with time as $\exp(\Gamma(k)t)$. In order to calculate the position of the  center of mass of the wavepacket $\langle k \rangle = \int k p(k) dk$ and its root mean square width $\Delta k=\sqrt{\langle k^2 \rangle - \langle k \rangle ^2}$, we use the normalized probability density accounting for the overall growth or decay of the wavepacket:
\begin{equation}
p(k)=\frac{|\psi(k)|^2\exp(\Gamma(k)t)}{\int|\psi(k)|^2\exp(\Gamma(k)t)\,dk}   
\end{equation}
Keeping the leading order terms, we obtain that the root mean square width of the wavepacket behaves as
\begin{equation}
    \Delta k(t)=\sqrt{\sigma_0^2-4\sigma_0\left(\log\sqrt{\frac{\sigma_0}{\sigma_\gamma}}-1\right)\sigma_\gamma \gamma t}
\end{equation}
where $\sigma_\gamma \gamma=\lambda U'^2/\alpha$,
and the root mean square width in real space $\Delta r$ grows as its inverse. 
The results are presented and discussed in the main text in the following form:
\begin{equation}
    \Delta r=\frac{A}{\sqrt{1-B\left(\frac{I_2}{I_1}\right)^2}}\text{.}
\end{equation}
Here, $A$ is the size of the probe without expansion (size of a single lattice site), whereas $B\approx 4\gamma t\sigma_\gamma/\sigma_0$ (with $t$ the effective propagation time). Since $\sigma_\gamma\ll\sigma_0$ (losses due to the quasicrystal lattice are localized in the reciprocal space), one needs $4\gamma t\gg 1$ for $B$ to be of the order of $1$: it means that the non-Hermiticity needs to be sufficiently strong to produce non-negligible effects during the propagation through the vapor cell.

\bibliographystyle{Science}
\bibliography{biblio}

\begin{thebibliography}{10}

\bibitem{Levine1984}
D.~Levine, P.~J. Steinhardt, {\it Phys. Rev. Lett.\/} {\bf 53}, 2477 (1984).

\bibitem{Shechtman1984}
D.~Shechtman, I.~Blech, D.~Gratias, J.~W. Cahn, {\it Phys. Rev. Lett.\/} {\bf 53}, 1951 (1984).

\bibitem{Ishimasa1985}
T.~Ishimasa, H.-U. Nissen, Y.~Fukano, {\it Phys. Rev. Lett.\/} {\bf 55}, 511 (1985).

\bibitem{yang1987description}
Q.~Yang, W.~Wei, {\it Phys. Rev. Lett.\/} {\bf 58}, 1020 (1987).

\bibitem{Gahler1988}
F.~Gahler, {\it Quasicrystalline materials : Proceedings of the I.L.L. / Codest Workshop, Grenoble, 21 - 25 March 1988\/}, C.~Janot, ed. (World Scientific, Singapore, 1988), chap.~7, pp. 272--284.

\bibitem{niizeki1987two}
N.~Niizeki, H.~Mitani, {\it J. Phys. A: Math. Gen.\/} {\bf 20}, L405 (1987).

\bibitem{Zhang2001}
X.~Zhang, Z.-Q. Zhang, C.~T. Chan, {\it Phys. Rev. B\/} {\bf 63}, 081105 (2001).

\bibitem{ahn2018dirac}
S.~J. Ahn, {\it et~al.\/}, {\it Science\/} {\bf 361}, 782 (2018).

\bibitem{crosse2021trigonal}
J.~Crosse, P.~Moon, {\it Sci. Rep.\/} {\bf 11}, 11548 (2021).

\bibitem{andrei2020graphene}
E.~Y. Andrei, A.~H. MacDonald, {\it Nat. Mater.\/} {\bf 19}, 1265 (2020).

\bibitem{Tarnopolsky2019}
G.~Tarnopolsky, A.~J. Kruchkov, A.~Vishwanath, {\it Phys. Rev. Lett.\/} {\bf 122}, 106405 (2019).

\bibitem{Moon2019}
P.~Moon, M.~Koshino, Y.-W. Son, {\it Phys. Rev. B\/} {\bf 99}, 165430 (2019).

\bibitem{Huang2019}
B.~Huang, W.~V. Liu, {\it Phys. Rev. B\/} {\bf 100}, 144202 (2019).

\bibitem{Hayashida2007}
K.~Hayashida, T.~Dotera, A.~Takano, Y.~Matsushita, {\it Phys. Rev. Lett.\/} {\bf 98}, 195502 (2007).

\bibitem{zhang2012dodecagonal}
J.~Zhang, F.~S. Bates, {\it J. Am. Chem. Soc.\/} {\bf 134}, 7636 (2012).

\bibitem{gillard2016dodecagonal}
T.~M. Gillard, S.~Lee, F.~S. Bates, {\it Proc. Natl. Acad. Sci.\/} {\bf 113}, 5167 (2016).

\bibitem{jayaraman2021dodecagonal}
A.~Jayaraman, C.~M. Baez-Cotto, T.~J. Mann, M.~K. Mahanthappa, {\it Proc. Natl. Acad. Sci.\/} {\bf 118}, e2101598118 (2021).

\bibitem{fischer2011colloidal}
S.~Fischer, {\it et~al.\/}, {\it Proc. Natl. Acad. Sci.\/} {\bf 108}, 1810 (2011).

\bibitem{xiao2012dodecagonal}
C.~Xiao, N.~Fujita, K.~Miyasaka, Y.~Sakamoto, O.~Terasaki, {\it Nature\/} {\bf 487}, 349 (2012).

\bibitem{Kraus2013}
Y.~E. Kraus, Z.~Ringel, O.~Zilberberg, {\it Phys. Rev. Lett.\/} {\bf 111}, 226401 (2013).

\bibitem{Tran2015}
D.-T. Tran, A.~Dauphin, N.~Goldman, P.~Gaspard, {\it Phys. Rev. B\/} {\bf 91}, 085125 (2015).

\bibitem{Hua2020}
C.-B. Hua, R.~Chen, B.~Zhou, D.-H. Xu, {\it Phys. Rev. B\/} {\bf 102}, 241102 (2020).

\bibitem{chan1998photonic}
Y.~Chan, C.~T. Chan, Z.~Liu, {\it Phys. Rev. Lett.\/} {\bf 80}, 956 (1998).

\bibitem{kaliteevski2000two}
M.~Kaliteevski, {\it et~al.\/}, {\it Nanotechnology\/} {\bf 11}, 274 (2000).

\bibitem{zoorob2000complete}
M.~Zoorob, M.~Charlton, G.~Parker, J.~Baumberg, M.~Netti, {\it Nature\/} {\bf 404}, 740 (2000).

\bibitem{Feng2005}
Z.~Feng, {\it et~al.\/}, {\it Phys. Rev. Lett.\/} {\bf 94}, 247402 (2005).

\bibitem{man2005experimental}
W.~Man, M.~Megens, P.~J. Steinhardt, P.~M. Chaikin, {\it Nature\/} {\bf 436}, 993 (2005).

\bibitem{gauthier2005photonic}
R.~C. Gauthier, K.~Mnaymneh, {\it Opt. Express\/} {\bf 13}, 1985 (2005).

\bibitem{nozaki2006lasing}
K.~Nozaki, T.~Baba, {\it Jpn. J. Appl. Phys.\/} {\bf 45}, 6087 (2006).

\bibitem{Ren2018}
J.~Ren, X.~Sun, S.~Wang, {\it Opt. Laser Technol.\/} {\bf 101}, 42 (2018).

\bibitem{Xi2019}
X.~Xi, X.~Sun, {\it Superlattices Microstruct.\/} {\bf 129}, 247 (2019).

\bibitem{aubry1980analyticity}
S.~Aubry, G.~Andr{\'e}, {\it Ann. Israel Phys. Soc\/} {\bf 3}, 18 (1980).

\bibitem{Lahini2009}
Y.~Lahini, {\it et~al.\/}, {\it Phys. Rev. Lett.\/} {\bf 103}, 013901 (2009).

\bibitem{Ganeshan2013}
S.~Ganeshan, K.~Sun, S.~Das~Sarma, {\it Phys. Rev. Lett.\/} {\bf 110}, 180403 (2013).

\bibitem{dominguez2019aubry}
G.~Dom{\'\i}nguez-Castro, R.~Paredes, {\it Eur. J. Phys.\/} {\bf 40}, 045403 (2019).

\bibitem{Longhi2019}
S.~Longhi, {\it Phys. Rev. Lett.\/} {\bf 122}, 237601 (2019).

\bibitem{simon1982almost}
B.~Simon, {\it Adv. Appl. Math.\/} {\bf 3}, 463 (1982).

\bibitem{bellissard1992gap}
J.~Bellissard, A.~Bovier, J.-M. Ghez, {\it Rev. Math. Phys.\/} {\bf 4}, 1 (1992).

\bibitem{gambaudo2014brillouin}
J.-M. Gambaudo, P.~Vignolo, {\it New J. Phys\/} {\bf 16}, 043013 (2014).

\bibitem{Tanese2014fractal}
D.~Tanese, {\it et~al.\/}, {\it Phys. Rev. Lett.\/} {\bf 112}, 146404 (2014).

\bibitem{roati2008anderson}
G.~Roati, {\it et~al.\/}, {\it Nature\/} {\bf 453}, 895 (2008).

\bibitem{goblot2020emergence}
V.~Goblot, {\it et~al.\/}, {\it Nat. Phys.\/} {\bf 16}, 832 (2020).

\bibitem{ueda1987energy}
K.~Ueda, H.~Tsunetsugu, {\it Phys. Rev. Lett.\/} {\bf 58}, 1272 (1987).

\bibitem{Szabo2020}
A.~Szab\'o, U.~Schneider, {\it Phys. Rev. B\/} {\bf 101}, 014205 (2020).

\bibitem{Gautier2021}
R.~Gautier, H.~Yao, L.~Sanchez-Palencia, {\it Phys. Rev. Lett.\/} {\bf 126}, 110401 (2021).

\bibitem{zhu2023localization}
Z.~Zhu, S.~Yu, D.~Johnstone, L.~Sanchez-Palencia, {\it arXiv\/} p. 2307.09527 (2023).

\bibitem{damanik2011spectral}
D.~Damanik, A.~Gorodetski, {\it Commun. Math. Phys.\/} {\bf 305}, 221 (2011).

\bibitem{sutHo1989singular}
A.~S{\"u}t{\H{o}}, {\it J. Stat. Phys.\/} {\bf 56}, 525 (1989).

\bibitem{he1989eigenvalue}
S.~He, J.~Maynard, {\it Phys. Rev. Lett.\/} {\bf 62}, 1888 (1989).

\bibitem{vardeny2013optics}
Z.~V. Vardeny, A.~Nahata, A.~Agrawal, {\it Nat. Photonics\/} {\bf 7}, 177 (2013).

\bibitem{levi2011disorder}
L.~Levi, {\it et~al.\/}, {\it Science\/} {\bf 332}, 1541 (2011).

\bibitem{Ozdemir2019}
{\c{S}}.~{\"O}zdemir, S.~Rotter, F.~Nori, L.~Yang, {\it Nat. Mater.\/} {\bf 18}, 783 (2019).

\bibitem{Lin2022}
Q.~Lin, {\it et~al.\/}, {\it Phys. Rev. Lett.\/} {\bf 129}, 113601 (2022).

\bibitem{Zeng2020}
Q.-B. Zeng, Y.-B. Yang, Y.~Xu, {\it Phys. Rev. B\/} {\bf 101}, 020201 (2020).

\bibitem{Li2022}
T.~Li, Y.-S. Zhang, W.~Yi, {\it Phys. Rev. B\/} {\bf 105}, 125111 (2022).

\bibitem{xu2022exact}
Z.-H. Xu, X.~Xia, S.~Chen, {\it Sci. China: Phys., Mech. Astron.\/} {\bf 65}, 227211 (2022).

\bibitem{sarnak1982spectral}
P.~Sarnak, {\it Commun. Math. Phys.\/} {\bf 84}, 377 (1982).

\bibitem{PhysRevA.51.576}
J.~Gea-Banacloche, Y.-q. Li, S.-z. Jin, M.~Xiao, {\it Phys. Rev. A\/} {\bf 51}, 576 (1995).

\bibitem{Zhang2019}
Z.~Zhang, {\it et~al.\/}, {\it Phys. Rev. Lett.\/} {\bf 122}, 233905 (2019).

\bibitem{zhang2020spin}
Z.~Zhang, {\it et~al.\/}, {\it Optica\/} {\bf 7}, 455 (2020).

\bibitem{zhang2022imaging}
Z.~Zhang, {\it et~al.\/}, {\it Photonics Res.\/} {\bf 10}, 958 (2022).

\bibitem{Zhang2020}
Z.~Zhang, {\it et~al.\/}, {\it Nat. Commun.\/} {\bf 11}, 1902 (2020).

\bibitem{PhysRevLett.131.013802}
Y.~Feng, {\it et~al.\/}, {\it Phys. Rev. Lett.\/} {\bf 131}, 013802 (2023).

\bibitem{miyai2006lasers}
E.~Miyai, {\it et~al.\/}, {\it Nature\/} {\bf 441}, 946 (2006).

\bibitem{perez2017demand}
B.~Perez-Garcia, C.~L{\'o}pez-Mariscal, R.~I. Hernandez-Aranda, J.~C. Guti{\'e}rrez-Vega, {\it Appl. Opt.\/} {\bf 56}, 6967 (2017).

\bibitem{fu2020optical}
Q.~Fu, {\it et~al.\/}, {\it Nature Photonics\/} {\bf 14}, 663 (2020).

\bibitem{jin1999band}
C.~Jin, {\it et~al.\/}, {\it Appl. Phys. Lett.\/} {\bf 75}, 1848 (1999).

\bibitem{di2008parametric}
E.~Di~Gennaro, {\it et~al.\/}, {\it Photonics Nanostruct.\/} {\bf 6}, 60 (2008).

\end{thebibliography}

\section*{Acknowledgments}

\textbf{Funding:} This work was supported by National Natural Science Foundation of China (No.51888103, No.62022066, and No.12074306) and European Union's Horizon 2020 program, through a FET Open research and innovation action under the grant agreement No. 964770 (TopoLight). We also acknowledge the support  of the ANR Labex Ganex (ANR-11-LABX-0014), and of the ANR program "Investissements d'Avenir" through the IDEX-ISITE initiative 16-IDEX-0001 (CAP 20-25).

\textbf{Author contributions:} \\
Conceptualization: ZZ, DS\\
Methodology: ZZ, DS\\
Investigation: ZZ, IS, DS, SL, JY\\
Formal analysis: IS, DS\\
Visualization: ZZ, IS, GM, DS, SL, JY\\
Supervision: ZZ, GM, DS\\
Resources: ZZ, GM, DS, ML, YZ, MX\\
Writing—original draft: DS\\
Writing—review and editing: ZZ, IS, GM, DS, YH.

\textbf{Competing interests:} The authors declare no competing interests. 

\textbf{Data and materials availability:} All data needed to evaluate the conclusions in the paper are present in the paper and/or the Supplementary Materials. Extended data generated in this study are available in the Open Science Framework (OSF) repository:

\verb|https://osf.io/2kdmc/?view_only=8b22bdaf561240c399c4c45a6af51d6e|

\end{document}